\documentclass[aps,prd,preprint,floatfix,nofootinbib,groupedaddress,showpacs]
{revtex4}
\usepackage{graphicx}
\usepackage{amsmath}
\usepackage{latexsym}
\usepackage{here}
\usepackage{array}
\usepackage{dcolumn}% Align table columns on decimal point
\usepackage{bm}% bold math
\def\lsim{\mathrel{\rlap{\raise 2.5pt \hbox{$<$}}\lower 2.5pt
\hbox{$\sim$}}}

\begin{document}
%\preprint{}
\title{On the branching ratio of the ``second class''
$\tau\to\eta^\prime\pi\nu_\tau$ decay}
%\thanks{}

\author{N. Paver}
\email[]{nello.paver@ts.infn.it}
%\thanks{}
\affiliation{Department of Physics, University of Trieste, 34100 Trieste, Italy \\
and \\
INFN-Sezione di
Trieste, 34100 Trieste, Italy}

\author{Riazuddin}
\email[]{riazuddin@ncp.edu.pk}
%\thanks{}
\affiliation{National Centre for Physics,\\ Quaid-i-Azam University Campus,\\
Islamabad, Pakistan}

\date{\today}

\begin{abstract}
We present an assessment of the standard model expectations  for the branching ratio of the isotopic spin and G-parity violating decay $\tau\to\eta^\prime\pi\nu_\tau$. This  estimate is based on a vector and scalar meson dominance parametrization of the relevant form factors, that explicitly accounts for $\pi^0-\eta-\eta^\prime$ mixing. The numerical results obtained in this framework indicate a branching ratio one order of magnitude (or more) below the current experimental limit, and suggest the possibility of evidencing some novel interaction in high statistics studies of this decay.
\end{abstract}

\pacs{13.35.Dx, 12.40.Vv}

\maketitle
%%%%%%%%%%%%%%%%%%%%%%%%%%%%%%%%%%%%%%%%%%%%%%%%%%%%%%%%%%%%%
The semileptonic transition $\tau\to\eta^\prime\pi\nu_\tau$ belongs to the category of the so-called ``second class current'' decays \cite{Weinberg:1958ut}. This kind of processes can in the standard model occur only {\it via} an isotopic spin and G-parity violation mechanism and, consequently, are  strongly suppressed by the small $d-u$ current quark mass difference. The present experimental limit,   
${{\rm BR}(\tau\to\eta^\prime\pi\nu_\tau)<7.2\times 10^{-6}}$~\cite{Aubert:2008nj}, is rather constraining. Nevertheless, depending on the actual smallness of the standard model branching ratio, this upper limit might still be high enough to allow some room available for revealing  ``nonstandard'' effects in high statistics studies of this decay, for example, at a Super B factory~\cite{O'Leary:2010af}. This has recently been emphasized with regard to $\tau\to\eta\pi\nu_\tau$ in Refs. 
\cite{Nussinov:2008gx,Paver:2010mz}, and earlier in Refs. \cite{Meurice:1987pp,Bramon:1987zb}, and should be the case also of the decay of interest here. 
\par 
While there are several theoretical estimates of ${\tau\to\eta\pi\nu_\tau}$, based either on the direct application of $\rho(770)$ and $a_0(980)$ pole dominance \cite{Nussinov:2008gx,Pich:1987qq} or on current algebra and chiral perturbation theory \cite{Neufeld:1994eg,Tisserant:1982fc}, recent attempts to estimate the branching ratio for 
$\tau\to\eta^\prime\pi\nu_\tau$ appeared in the literature are, to our knowledge, much fewer \cite{Nussinov:2009sn}. In this note, we extend to this process the method used for the calculation of $\tau\to\eta\pi\nu_\tau$ in Ref.~\cite{Paver:2010mz}. This 
approach consists of a parametrization of the spin-1 and spin-0 relevant form factors in the timelike region, in terms of $\rho(770),\rho^\prime\equiv\rho(1450)$ and $a_0(980),a_0^\prime\equiv a_0(1450)$ exchanges, respectively, with coupling constants either theoretically estimated within the quark model or limited from the available phenomenology, and constrained by the value of the isotopic spin violating $\pi^0-\eta$ mixing parameter earlier evaluated from chiral symmetry breaking. 
\par 
Indeed, in this regard, the scale of the transition amplitude for $\tau\to\eta\pi\nu_\tau$ is set by a 
$\pi^0-\eta$ mixing parameter $\epsilon_{\eta\pi}$ calculable in chiral perturbation theory to leading and next-to-leading order, see Refs.~\cite{Gasser:1984ux,Ecker:1999kr} and the application in Ref.~\cite{Neufeld:1994eg}. The scale of 
$\tau\to\eta^\prime\pi\nu_\tau$ can be set by a 
$\pi^0-\eta-\eta^\prime$ mixing parameter $\epsilon_{\eta^\prime\pi}$ that, the $\eta^\prime$ not being a Nambu-Goldstone boson except for $N_c\to
\infty$, can phenomenologically be estimated in a chiral symmetry breaking scheme supplemented by the determination of the $\eta-\eta^\prime$ mixing angle \cite{Leutwyler:1996tz}. The other, obvious, difference between the two processes is a kinematical one, namely, in the case of $\tau\to\eta^\prime\pi\nu_\tau$ the $1^-$ and $0^+$ exchanged ground states are either well-below or just around (if the width is taken into account) the threshold of the decay phase space.   
\par
With $V_\mu={\bar u}\gamma_\mu d$ the weak vector
current, the hadronic matrix element for
$\tau\to\eta^\prime\pi\nu_\tau$ can be decomposed into
spin-1 and spin-0 exchange form factors,
respectively, as
\begin{equation}
\label{formfactors}
\langle\pi^+(k)\eta^\prime(p)\vert V_\mu \vert 0\rangle =
-\sqrt{2}\left[f_1(t)\left(\left(p-k\right)_\mu -
\frac{M^2_{\eta^\prime}-M^2_\pi}{t}q_\mu\right) +
f_0(t)\frac{M^2_{\eta^\prime}-M^2_\pi}{t}q_\mu\right].
\end{equation}
Here, $q=p+k$ and $t=q^2$ is the invariant mass
squared of the emitted $\eta^\prime\pi$ pair. 
%Finiteness of the matrix element at $t=0$ implies the %condition$f_1(0)=f_0(0)$.
\par 
With $t_0=(M_{\eta^\prime}+M_\pi)^2$ and
$\lambda(x,y,z)=x^2+y^2+z^2 -2(xy+yz+zx)$, the partial 
width is:
\begin{eqnarray}
\label{decayrate}
\Gamma(\tau^+\to\eta^\prime\pi^+\nu_\tau) &=&
\frac{G_F^2\vert V_{ud}\vert^2}{384\pi^3M_\tau^3}
\int_{t_0}^{M_\tau^2}\frac{dt}{t^3}
\lambda^{1/2}(t,M^2_{\eta^\prime},M^2_\pi)(M_\tau^2-t)^2
\nonumber \\
&\times& \left[\vert f_1(t)\vert^2
\left((2t+M_\tau^2)\lambda(t,M^2_{\eta^\prime},M_\pi^2)\right) +
\vert f_0(t)\vert^2 3M_\tau^2(M^2_{\eta^\prime}-M_\pi^2)^2\right]
\end{eqnarray}
\par 
For the spin-1 form factor we assume, along the lines of Refs.~\cite{Bruch:2004py,Dominguez:2001zu}, the unsubtracted linear combination of $\rho$ and $\rho^\prime$ polar forms 
\begin{equation}
\label{spinone}
f_1(t)={\displaystyle \frac{f_\rho g_{\rho{\eta^\prime}\pi}}{M_\rho^2} 
\left[\frac{M_\rho^2}{M_\rho^2-t} +
\beta_\rho\frac{M_{\rho^\prime}^2}{M_{\rho^\prime}^2-t-
iM_{\rho^\prime}\Gamma_{\rho^\prime}(t)}\right]}.
\end{equation}
Here, denoting by $M$ and $\Gamma$ the 
mass and total width of a spin-L resonance \cite{Kuhn:1990ad}: 
\begin{equation}
\label{width}
\Gamma(t)=\theta(t-t_0) \frac{M}{\sqrt t}
\left(\frac{q(t)}{q(M^2)}\right)^{2{\rm L}+1} \Gamma, 
\end{equation}
where in our case $q$ is the momentum in the $\eta^\prime\pi$ c.m. frame and $L=1,0$ for spin-1 and 0, respectively. Notice that, in Eq.~(\ref{spinone}), we have omitted the 
$\rho$ width, this pole is below the threshold $t_0$ and  the numerical results are almost insensitive to this approximation. 
In Eq.~(\ref{spinone}), $f_\rho$ and ${\displaystyle{g_{\rho{\eta^\prime}\pi}}}$ are the vector 
meson couplings to $e^+e^-$ and to $\eta^\prime\pi$, respectively, and the coefficient 
${\beta_\rho=\displaystyle (M_\rho/M_{{\rho^\prime}})^2\times (f_{{\rho^\prime}} 
g_{{\rho^\prime}{\eta^\prime}\pi}/f_\rho g_{\rho{\eta^\prime}\pi})}$ parametrizes the contribution of the radial excitation $\rho^\prime$. In some sense, 
Eq.~(\ref{spinone}) resembles the modification of the 
$\rho$ propagator introduced in Ref.~\cite{Dumm:2009va}.
\par 
We incorporate isotopic spin violation through 
the $\eta^\prime\leftrightarrow\pi^0$ transition 
${\langle\pi^0\vert{\rm H}^\prime\vert\eta^\prime\rangle}$.  In the quark model, neglecting isospin breaking of electromagnetic origin, ${\rm H}^\prime=-\Delta m S^3$ with  $\Delta m=m_d-m_u$ and the scalar densities 
$S^i={\bar q}\frac{\lambda^i}{2}q$ ($\lambda$'s are Gell-Mann matrices). To leading order in $\Delta m$, the 
${\displaystyle(\rho{\eta^\prime}\pi)}$ and 
$({\rho^\prime}{\eta^\prime}\pi)$ trilinear couplings will be assumed to be proportional to the $(\rho\pi\pi)$ and 
$({\rho^\prime}\pi\pi$) ones, through the mixing parameter 
\begin{equation}
\label{epsilonprime}
{\displaystyle\epsilon_{{\eta^\prime}\pi}}=
{\displaystyle \frac{\langle\pi^0\vert{\rm H}^\prime\vert\eta^\prime\rangle}
{M^2_\pi-M^2_{\eta^\prime}}},
\end{equation} 
so that 
\begin{equation}
\label{vectorgis}
g_{\rho{\eta^\prime}\pi}=
\epsilon_{{\eta^\prime}\pi}
g_{\rho\pi\pi};
\qquad\quad 
g_{{\rho^\prime}{\eta^\prime}\pi}=
\epsilon_{{\eta^\prime}\pi}
g_{{\rho^\prime}\pi\pi}.
\end{equation}
\par 
In the soft-pion limit, by applying current algebra:
\begin{equation}
\label{hprime}
\langle\pi^0\vert{\rm H}^\prime\vert\eta^\prime\rangle=
-\frac{\Delta m}{\sqrt 3 F_\pi}
\langle 0\vert\sqrt 2 P_0 + P_8\vert\eta^\prime\rangle, 
\end{equation}
with $F_\pi$ the pion decay 
constant and $P^i$ the pseudoscalar densities 
$P^i={\bar q}\frac{\lambda^i}{2}\gamma_5q$. Similar relations as 
in Eqs.~(\ref{epsilonprime}) and (\ref{hprime}) can be derived also for the case of the $\eta$. 
\par 
For the $\eta$ and $\eta^\prime$ states we adopt the single mixing angle scheme in terms of the pure octet and singlet pseudoscalars, 
\begin{equation}
\label{mixingangle}
\vert\eta\rangle=\cos\theta\vert\eta_8\rangle-
\sin\theta\vert\eta_0\rangle; 
\qquad 
\vert\eta^\prime\rangle=\sin\theta\vert\eta_8\rangle+
\cos\theta\vert\eta_0\rangle.
\end{equation} 
By relating the $P^i$ in Eq.~(\ref{hprime}) to 
the divergences of the octet and singlet axial currents 
($U(1)$ anomaly included), in the $SU(2)\times SU(2)$ limit, 
after some algebra one obtains the relation
\begin{equation}
\langle 0\vert\sqrt 2 P_0+p_8\vert\eta^\prime\rangle=
\frac{3}{4}F_8\frac{\sin\theta}{m_s}\bigl[\cos^2\theta M^2_\eta+\sin^2\theta M^2_{\eta^\prime}-
2\cos^2\theta\left(M^2_{\eta^\prime}-M^2_\eta\right)\bigr].
\label{p0+p8}
\end{equation} 
We take for the mixing angle the value $\theta=-20^\circ$ 
\cite{Gilman:1987ax}. Values of $\theta$ in the range, say, 
$[-15^\circ, -20^\circ]$ have been obtained 
recently~\cite{Pham:2010sr}, depending on the physical process and on the model, and also two-angles $\eta-\eta^\prime$ mixing schemes have been proposed~\cite{Escribano:2005qq}. On the other hand, the dependence of Eq.~(\ref{p0+p8}) on $\theta$ is rather mild. By combining Eqs.~(\ref{epsilonprime})-(\ref{p0+p8}), 
assuming the quark mass ratios $m_u/m_d\simeq 0.55$ and  
$m_s/m_d\simeq 18.9$~\cite{Leutwyler:1996qg}, and 
the pseudoscalar decay constants $F_8\simeq F_\pi$, we find the approximate value 
${\displaystyle\epsilon_{{\eta^\prime}\pi}}
\simeq 3\times 10^{-3}$. To encompass the other 
determinations~\cite{Leutwyler:1996tz,Kroll:2005sd,Chao:1989yp} and 
somehow account for theoretical uncertainties, we in 
the sequel will allow for $\epsilon_{{\eta^\prime}\pi}$ the 
range of values
\begin{equation}
\label{epsilonprimet} 
\epsilon_{{\eta^\prime}\pi}=(3\pm 1)\times 10^{-3}. 
\end{equation}
Notice that, with these values, and 
$g_{\rho\pi\pi}=6$ from the experimental $\rho$-width, the determination of the $(\rho{\eta^\prime}\pi)$ 
coupling constant from Eq.~(\ref{vectorgis}) is somewhat  smaller than obtained in  
Ref.~\cite{Genz:1994zp}, and falls well-below the upper limit derived in Ref.~\cite{Nussinov:2009sn}. 
\par
As regards the estimate of the parameter $\beta_\rho$ in 
Eq.~(\ref{spinone}), a calculation within the constituent 
quark model, using the ratio of $\rho$ and $\rho^\prime$ wave functions at the origin~\cite{Eichten:1979ms}, indicates  $f_{\rho^\prime}/f_\rho\simeq 1.1$. Moreover, 
an assessment (actually, an upper limit) of $g_{{\rho^\prime}\pi\pi}$ can be obtained by identifying 
$\Gamma(\rho^\prime\to\pi\pi)$ to $\Gamma_{\rho^\prime}\simeq 0.4$~GeV. This gives for $\beta_\rho$ the upper limit: 
$\vert\beta_\rho\vert\leq 0.18$. We finally complete the 
numerical input needed for the parametrization (\ref{spinone}) by 
taking, from experimental data on $\rho$ decays, 
$f_\rho g_{\rho\pi\pi}/M^2_{\rho}\simeq 1.2$~\cite{pdg}. 
\par 
We turn to the spin-0 exchange form factor $f_0(t)$, 
determined by the matrix element of the divergence of the vector current in (\ref{formfactors}), through the relation 
exhibiting the isotopic spin violation proportional to $\Delta m$: \begin{equation}
\label{divergence}
\langle\pi^+(k){\eta^\prime}(p)\vert\ i \partial^\mu V_\mu\vert 0\rangle
=(m_d-m_u)\langle\pi^+(k){\eta^\prime}(p)\vert S^{(1+i2)}\vert 0\rangle
=\sqrt{2}\left(M^2_{\eta^\prime}-M^2_\pi\right)f_0(t).
\end{equation}
We assume a similar parametrization as in Eq.~(\ref{spinone}), dominated by $a_0$ and $a_0^\prime$ poles: 
\begin{equation}
\label{spinzero}
f_0(t)=\frac{F_{a_0}g_{a_0{\eta^\prime}\pi}}{M^2_{\eta^\prime}-M^2_\pi}
\left[\frac{M^2_{a_0}}{M^2_{a_0}-t-iM_{a_0}\Gamma_{a_0}} + \beta_a
\frac{M^2_{a_0^\prime}}{M^2_{a_0^\prime}-t-iM_{a_0^\prime}\Gamma_{a_0^\prime}(t)}\right].
\end{equation}
Here, the factor $\beta_a$ multiplying the $a_0^\prime$ pole is defined by the ratio: 
$\beta_a=F_{a_0^\prime}g_{{a_0^\prime}{\eta^\prime}\pi}/F_{a_0}g_{a_0{\eta^\prime}\pi}$, with 
$g_{a_0{\eta^\prime}\pi}$ and $g_{{a_0^\prime}{\eta^\prime}\pi}$
the trilinear $(a_0{\eta^\prime}\pi)$ and 
$({a_0^\prime}{\eta^\prime}\pi)$ coupling constants, 
respectively, and 
\begin{equation}
\label{fazero}
-\sqrt{2}F_{a_0}M_{a_0}^2=
\langle a_0(q)\vert i\partial^\mu V_\mu\vert 0\rangle,
\qquad
-\sqrt{2}F_{a_0^\prime}M_{a_0^\prime}^2 =
\langle a_0^\prime(q)\vert i\partial^\mu V_\mu\vert 0\rangle.
\end{equation}
In Eq.~(\ref{spinzero}), we neglect the $t$-variation of 
$\Gamma_{a_0}$ in order not to generate, with $M_{a_0}$ so close to the threshold $t_0$, spurious imaginary parts.
\par
In a $U(3)$ scheme with $\eta-\eta^\prime$ mixing according to (\ref{mixingangle}), one obtains for $\theta=-20^\circ$:
\begin{equation}
\label{getaprime}
\frac{g_{a_0{\eta^\prime}\pi}}{g_{a_0\eta\pi}}=
\frac{g_{{a_0^\prime}{\eta^\prime}\pi}}
{g_{{a_0^\prime}\eta\pi}}=
\frac{(\sqrt{2}\cos\theta+\sin\theta)}
{(\cos\theta-\sqrt{2}\sin\theta)}=0.70.
\end{equation}  
This relation is in agreement with the experimental ratio 
$\Gamma(a_0^\prime\to\eta^\prime\pi)/
\Gamma(a_0^\prime\to\eta\pi)\simeq 0.35$ \cite{pdg}, see also Ref.~\cite{Parganlija:2010fz}. Along the lines 
of Ref.~\cite{Paver:2010mz}, to exploit Eq.~(\ref{getaprime}) we adopt the experimental value 
$g_{a_0\eta\pi}\simeq 2.80\ {\rm GeV}$
accompanied by a total width
$\Gamma_{a_0}\simeq 100\ {\rm MeV}$ 
\cite{Ambrosino:2009py}. For the
$g_{a_0^\prime\eta\pi}$ coupling constant, we assume
that the $\eta\pi$, $\eta^\prime\pi$, $K{\bar K}$ and
$\omega\pi\pi$ decay channels of the $a_0^\prime$
saturate the total width
$\Gamma_{a_0^\prime}=265\ {\rm MeV}$.
Using the ratios between partial widths
$\Gamma(\eta^\prime\pi)/\Gamma(\eta\pi)$,
$\Gamma(K\bar K)/\Gamma(\eta\pi)$ and
$\Gamma(\omega\pi\pi)/\Gamma(\eta\pi)$ reported
in \cite{pdg}, one would find
$\Gamma(a_0^\prime\to\eta\pi)\simeq 20\ {\rm MeV}$,
and the consequent estimate
$g_{a_0^\prime\eta\pi}\simeq 1.32\ {\rm GeV}$. Notice that 
these values agree to a good extent with the predictions  recently derived in Ref.~\cite{GarciaRecio:2010ki} from an SU(6) breaking approach, and in Ref.~\cite{Wang:2008wc} using QCD sum rules.
\par
To estimate $\beta_a$, we further need the values of the 
the constants $F_{a_0}$ and $F_{a_0^\prime}$ in 
(\ref{fazero}), and we assume both $0^+$ scalars to be 
$p$-wave ${\bar q}q$ states. Similar to \cite{Nussinov:2009sn}, one might adopt the $SU(6)$ framework and relate, by means of current algebra equal-time commutators and single-particle saturation of the ensuing sum rules, $F_{a_0}$ to the analogous $1^+$ constant 
$F_{a_1}$. In the $SU(2)\times SU(2)$ limit, the 2nd 
Weinberg sum rule~\cite{Weinberg:1967kj} would then 
relate $F_{a_1}= F_\rho$. 
Using the same procedure for $F_{a_0^\prime}$, we would 
obtain the ratio \cite{Paver:2010mz}:
\begin{equation}
\label{su6}
\frac{F_{a_0^\prime}}{F_{a_0}}\simeq
\frac{M_{a_0}^2}{M_{a_0^\prime}^2} \
\frac{F_{\rho^\prime}}{F_\rho},
\end{equation}
with $F_{\rho^\prime}/F_\rho$ previously estimated, 
hence the estimate $\vert\beta_a\vert=0.23$. 
\par 
Anther possibility would be to use the QCD sum rules and 
local hadron duality relations $M_S^2(n)=(n+1)M_S^2$ and   
$F_S^2(n)=F_S^2/(n+1)$, where $S$ denotes $0^+$ scalar 
\cite{Kataev:2004ve,Gorishnii:1983zi}. For $n=1$, the 
former relation is in perfect agreement with the 
experimental $M_{a_0}$ and $M_{a_0^\prime}$; the latter 
one indicates $F_{a_0^\prime}/F_{a_0}=0.70$. Numerically, 
with this approach we get the estimate 
$\vert\beta_a\vert\simeq 0.33$, which we ultimately adopt 
for the final numerical calculations and, 
anyway, is compatible within our uncertainties to the
value obtained before.
\par
Finally, theoretical estimates of $F_{a_0}$ are available, 
and we take the values $F_{a_0}\simeq 1.3-1.6\ {\rm MeV}$ 
obtained from QCD sum rules \cite{Narison:2008nj}.
\par
Inserting in Eq.~(\ref{decayrate}) the 
parametrizations (\ref{spinone}) and (\ref{spinzero}) 
with the input coupling constants derived above, in 
particular with $\epsilon_{{\eta^\prime}\pi}$ as in 
(\ref{epsilonprimet}), $\vert\beta_\rho\vert\leq 0.18$,  
consistent with~\cite{Fujikawa:2008ma}, 
and $\vert\beta_a\vert\leq 0.33$, we obtain predictions 
for the decay rates from spin-1 and spin-0 exchanges. 
For any fixed pair of $\beta_\rho$, $\beta_a$ 
in the mentioned ranges, we fix the corresponding 
value of $\epsilon_{{\eta^\prime}\pi}$ lying 
in the range (\ref{epsilonprimet}), such that, as 
needed to cancel the kinematical singularity in 
Eq.~(\ref{formfactors}), the condition 
$f_1(0)=f_0(0)$ is numerically satisfied. From this 
procedure, we find the following intervals for the 
spin-1 and spin-0 exchanges:
\begin{equation}
\label{spinonepred}
{\rm BR}(\tau\to{\eta^\prime}\pi\nu_\tau)\vert_{{\rm spin}-1}
=1.4\times 10^{-9}-3.4\times 10^{-8},
\end{equation}
and
\begin{equation}
\label{spinzeropred}
{\rm BR}(\tau\to{\eta^\prime}\pi\nu_\tau)\vert_{{\rm spin}-0}= 
6.0\times 10^{-8}-1.8\times 10^{-7}.
\end{equation} 
With the current, limited, knowledge of the 
$\rho^\prime$ and $a_0^\prime$ properties, and 
the generous limits allowed to $\beta_\rho$ and 
$\beta_a$, we are not able to predict the branching 
ratio of $\tau\to\eta^\prime\pi\nu_\tau$, with a 
fully theoretical calculation, to a better accuracy. 
As one can see, Eqs.~(\ref{spinonepred}) and 
(\ref{spinzeropred}) clearly reproduce the general 
expectation of the phase space suppression of the 
spin-1 exchange vs. the spin-0 one. The upper values 
in these equations fall below the upper limit 
of the order of $10^{-6}$ presented in \cite{Nussinov:2009sn}, and   
should encourage high statistics studies of this 
process in the quest for ``non-standard" exchanges.

\newpage
%%%%%%%%%%%%%%%%%%%%%%%%%%%%%%%%%%%%%%%%%%%%%%%%%%%%%%%%%%%%%%%%%%%
%\section*{Acknowledgements}
%\medskip
\leftline{\bf Acknowledgments}
\par\noindent
This research has been partially supported by funds of the 
University of Trieste. One of the
authors (R) would like to thank the Abdus Salam ICTP
for hospitality.
%%%%%%%%%%%%%%%%%%%%%%%%%%%%%%%%%%%%%%%%%%%%%%%%%%%%%%%%%%%%%%%%%%%

\goodbreak
%%%%%%%%%%%%%%%%%%%%%%%%%%%%%%%%%%%%%%%%%%%%%%%%%%%%%%%%%%%%

\end{document}